\documentclass[10pt, twocolumn]{IEEEtran}

\usepackage{graphicx,amsmath,amssymb,cite}
\usepackage{subfigure}

\usepackage[dvips]{color}
\usepackage{float}
\ifCLASSINFOpdf
\else
\fi
\begin{document}
\title{Secure Precise Wireless Transmission with Random-Subcarrier-Selection-based Directional Modulation Transmit Antenna Array}

\author{Feng Shu, Xiaomin~Wu,~Jinsong Hu,~Riqing Chen,~and Jiangzhou Wang
\thanks{This work was supported in part by the National Natural Science Foundation of China (Nos. 61472190, and 61501238), the Open Research Fund of National Key Laboratory of Electromagnetic Environment, China Research Institute of Radiowave Propagation (No. 201500013).}
\thanks{~Feng~Shu,~Xiaomin~Wu,~Jinsong Hu are with School of Electronic and Optical Engineering, Nanjing University of Science and Technology, 210094, CHINA. E-mail:\{shufeng, xiaoming.wu\}@njust.edu.cn}
\thanks{Feng Shu and Riqing Chen are with the College of Computer and Information Sciences, Fujian Agriculture and Forestry University, Fuzhou 350002, China. E-mail: riqing.chen@fafu.edu.cn.}
\thanks{Feng~Shu is also with the National Key Laboratory of Electromagnetic Environment, China Research Institute of Radiowave Propagation, Qingdao 266107, China}
\thanks{Jiangzhou~Wang is  with  the School of Engineering and Digital Arts, University of Kent, Canterbury Kent CT2 7NT, United Kingdom}
}
\maketitle
\begin{abstract}
In this paper, a practical wireless transmission scheme is proposed to transmit confidential messages to the desired user securely and precisely by the joint use of multiple techniques including artificial noise (AN) projection, phase alignment (PA)/beamforming, and random subcarrier selection (RSCS) based on OFDM, and directional modulation (DM), namely RSCS-OFDM-DM. This RSCS-OFDM-DM scheme provides an extremely low-complexity structures for the transmitter and  desired receiver and makes the secure and precise wireless transmission realizable in practice. For illegal eavesdroppers, the receive power of confidential messages is so weak that their receivers cannot intercept these confidential messages successfully once it is corrupted by AN. In such a scheme, the design of phase alignment/beamforming vector and AN projection matrix depend intimately on the desired direction angle and distance. It is particularly noted that the use of RSCS leads to a significant outcome that the receive power of confidential messages mainly concentrates on the small neighboring region around the desired receiver and only small fraction of its power  leaks out to the remaining large broad regions. This concept is called secure precise transmission. The probability density function of real-time receive signal-to-interference-and-noise ratio (SINR) is derived.  Also, the average SINR and its tight upper bound are attained.  The approximate closed-form expression for average secrecy rate is derived by analyzing the first-null positions of SINR and clarifying the wiretap region. From simulation and analysis, it follows that the proposed scheme actually can achieve a secure and precise wireless transmission of confidential messages in line-of-propagation channel, and the derived theoretical formula of average secrecy rate  is verified to coincide with the exact one well for medium and large scale transmit antenna array or in the low and medium SNR regions.
\end{abstract}


\IEEEpeerreviewmaketitle

\section{Introduction}
In the last decade, secure physical-layer wireless transmission has increasingly become an extremely important research field in both academia and industry \cite{zhao,YAN1,zou1,zou2,YAN2,YAN3,Nusenu,HU3}. As a key-less physical layer secure transmit way, directional modulation (DM) is attracting an ever-growing interest in \cite{maha,kalantari,babakhani,daly,shiRF} and  has made substantial progression in many aspects by using antenna array with the help of aided artificial noise (AN) \cite{Ding,Hu1,Wu}. In \cite{daly,shiRF}, the authors propose an actively driven DM array of utilizing analog radio frequency (RF) phase shifters or switches, which exists some fundamental weaknesses such as requirement of high-speed RF switches and high complexity of design process. To overcome those disadvantages, instead of the RF frontend,  the DM is synthesized on the baseband by  the AN-aided orthogonal vector method in \cite{Ding}.  To enhance security,  the symbol-level precoder in \cite{kalantari} is proposed by using the concept of constructive interference in directional modulation with the goal of reducing the energy consumption at transmitter. Taking direction measurement error into account,  the authors in \cite{Hu1,Wu} propose two new robust DM synthesis methods for two different application scenarios: single-desired user and multi-user broadcasting by exploiting the statistical properties of direction measurement error. Compared to existing non-robust methods, the proposes robust methods can achieve an order-of-magnitude bit error rate (BER) performance improvement along desired drections. In general, the DM may achieve a high performance gain along the desired directions via beamforming and provide a secure transmission by seriously degrading the undesired directions by artificial noise(AN) projection.

However, the major disadvantage of DM is that its beamforming method only depends on direction angle and is independent of distance. Therefore, as an eavesdropper  moves within the main beam of the desired direction, it can readily intercept the confidential messages towards the desired direction. This will result in a serious secure problem.

To address such a problem, the authors in \cite{Liu,Hu2} propose a novel frequency diversity array model. The frequency increments are randomly (instead of linearly) assigned to all antenna-array elements, which is called a random frequency diverse array (RFDA) \cite{zhu}. In \cite{Hu2}, the authors propose a RFDA based directional modulation with aided AN. This scheme can achieve a secure precise wireless transmission. That is, the confidential beamforming vectors for desired users and AN projection  matrix for eavesdroppers rely heavily on both direction and distance. In other words, by making use of RFDA plus DM, the confidential messages can be securely and precisely transmitted towards a given position (direction and distance) with only a small fraction of confidential power radiating outside the small neighborhood around the desired position. Actually, the frequency diverse array has been widely investigated in RADAR field \cite{antonik,wang}. In \cite{Fusco}, the RFDA concept has also been employed to construct OFDM transmitter and achieved secure wireless communication in free space. However, as the number $N_T$ of transmit antennas tends to medium-scale and large-scale, the desired transmitter and receiver require a high-complexity transmit and receive structure. For example, at desired receiver, $N_T$ RF chains, or one RF chain with $N_T$ matched filter are required to coherently combine  $N_T$ parallel random independent subchannels of all $N_T$ frequency bands. This $N_T$-RF-chain structure will dramatically increase the circuit cost at both transmitter and receiver for medium-scale and large-scale directional modulation system. This motivates us propose a low-complexity structure of using random-subcarrier-selection (RSCS) array based on OFDM technique instead of RFDA with the help of AN projection and phase-alignment at transmitter. The proposed scheme can easily implement the DM synthesis at transmitter and coherent combining at receiver using IFFT/FFT operations,~respectively. Thus, this new structure can significantly reduce the circuit cost and make the secure precise communication feasible and applicable. Our main contributions are summarized as follows:

  $\bullet$ To address the problem of circuit complexity at receiver, we propose a new scheme of replacing random frequency diverse by random subcarrier selection based on OFDM and DM (RSCS-OFDM-DM). At desired receiver, the FFT/IFFT operation will reduce $N_T$ RF chains to only one. This simple structure will significantly save the circuit budget in the medium-scale and large-scale systems.

  $\bullet$  At transmitter,  phase alignment in frequency-domain is combined with RSCS to  constructively synthesize the $N_T$ frequency-shifted versions of transmit signal at desired receiver and destructively at eavesdroppers. The AN projection is used to further degrade the performance of eavesdroppers.

  $\bullet$  In the scenario that AN and channel noise have approximately the same receive power, the probability density function (PDF) of receive signal-to-interference-and-noise ratio (SINR) is first derived. Subsequently, the average SINR expression is presented by using the derived PDF.

  $\bullet$  The approximate closed-form expression for average secrecy rate is derived by analyzing the first-null position and clarifying the wiretap region. This theoretical formula is verified to be valid in medium and large scale transmit antenna array and almost independent of SNR. Additionally, in the low and medium SNR regions, the derived expression is shown to coincide with the exact one well.

  $\bullet$  By simulation and analysis,  we find that the receive confidential message power forms a high SINR peak around the desired position and shows a very small leakage to other wide regions. In other words, this gathers almost all confidential message power at a small neighbourhood around the desired position and at the same time leads to an extremely low receive SINR at eavesdropper positions by AN projection, RSCS and phase alignment at transmitter, which may successfully destroy the interception of eavesdroppers.

The remainder are organized as follows: Section II describes the system model of the proposed RSCS-OFDM-DM and proves its feasibility. The AN-aided and range-angle-dependent beamforming scheme for RSCS-OFDM-DM is presented in Section III. Here, we analyze the PDF of the receive SINR and its means. Also, we derive the theoretical formula for average secrecy rate by using the desired SINR and upper bound of eavesdropper SINR.  Section IV presents the simulation results and analysis. Finally, we make a conclusion in Section V.

Notations: throughout the paper, matrices, vectors, and scalars are denoted by letters of bold upper case, bold lower case, and lower case, respectively. Signs $(\cdot)^T$, and $(\cdot)^H$ denote transpose and conjugate transpose respectively. The operation $\|\cdot\|$ and $|\cdot|$ denote the norm of a vector and modulus of a complex number. The notation $\mathbb{E}\{\cdot\}$ refers to the expectation operation. Matrices $\textbf{I}_N$ denotes the $N\times N$ identity matrix and $\textbf{0}_{M\times N}$ denotes $M\times N$ matrix of all zeros.

\section{System model}
\begin{figure*}[htbp]
\centering
\includegraphics[width=15cm]{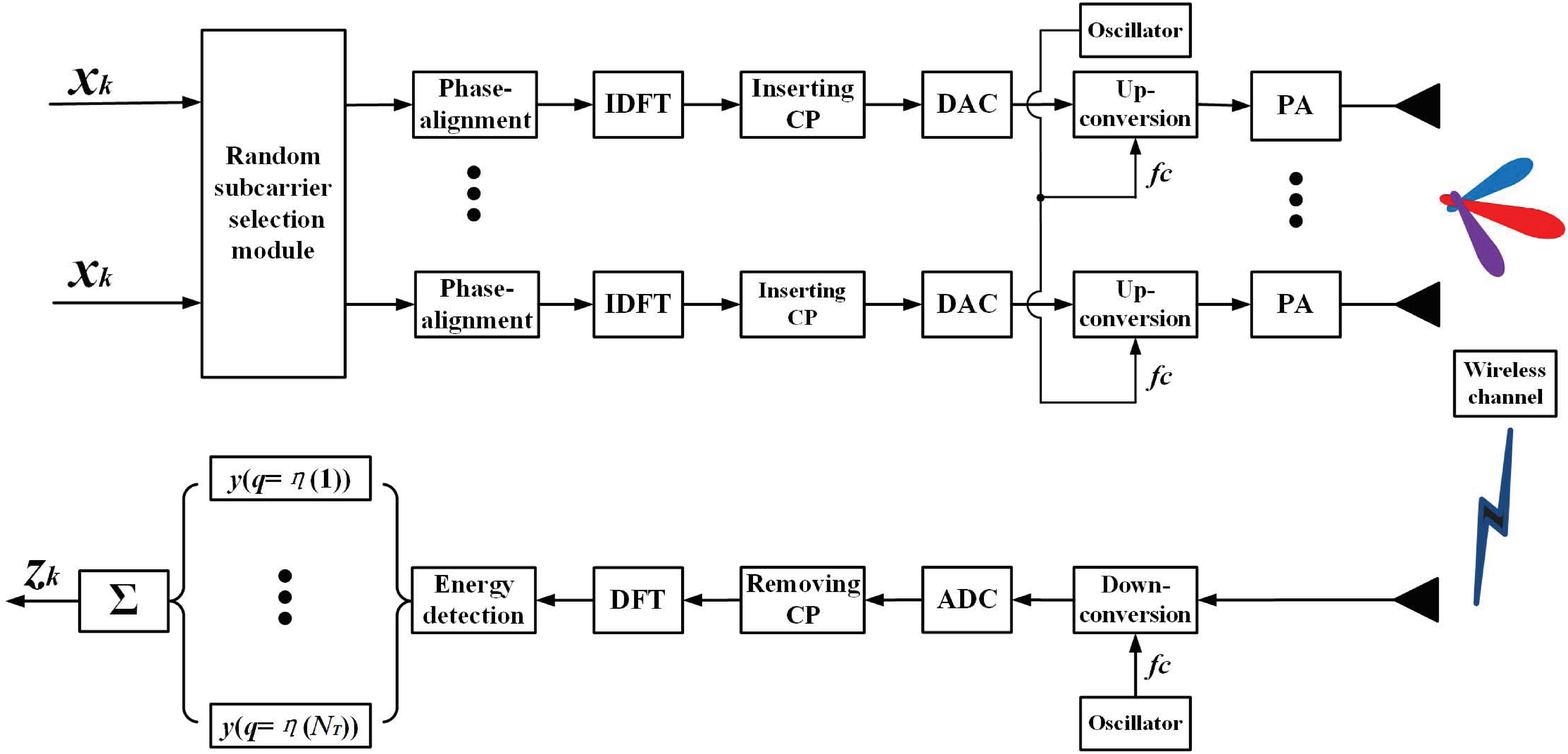}
\caption{Schematic diagram of RSCS-OFDM-DM communication system.}\label{fig1}
\end{figure*}

In this section, we propose a secure precise system structure as shown in Fig.~$\ref{fig1}$. This system consists of OFDM, random-subcarrier-selection (RSCS) and DM. In Fig.~$\ref{fig1}$, the RSCS-OFDM-DM communication system includes a legitimate transmitter equipped with one $N_T$-element linear antenna array,  a desired receiver, and several eavesdroppers, which lie at different positions from the desired position and is not shown in Fig.~$\ref{fig1}$, with each receiver having  single receive antenna.

The distinct frequency shifts of the same signal/symbol over different transmit antennas are  radiated as indicated in Fig.~$\ref{fig1}$. For the convenience of implementation of both transmitter and receiver, all frequency shifts are designed to be orthogonal with each other  in this paper. Thus, the subcarriers  in OFDM systems are a natural choice as shown in Fig.~2. Suppose there are $N$ subcarriers in our OFDM systems, the set of subcarriers is the following
\begin{align}
S_{sub}=\left\{f_m\rvert f_m=f_c+m\Delta f,~(m=0,1,\ldots,N-1)\right\},
\end{align}
where $f_c$ is the reference frequency, and $\Delta f$ is the subchannel bandwidth. We assume $N\Delta f\ll f_c$ in this paper. In our system, the total bandwidth is $B=N\Delta f$. The corresponding total subcarrier index set is defined as follows
\begin{align}
S_{N}=\{0,1,2,\ldots,N-1\}.
\end{align}
$N_T$ random subcarriers is chosen from $S_N$,  allocated to $N_T$ transmit antennas individually, and represented as $S_{N_T}$
\begin{align}
&~~~~~~S_{N_T}\subseteq S_N,~|S_{N_T}|=N_T.
\end{align}
Now, we define a chosen subcarrier index function $\eta(\bullet)$ as a mapping from  the set of transmit antenna indices $\left\{1,2,\cdots, N_T\right\}$ to  the chosen subcarrier set
\begin{align}
S_{N_T}=\left\{\eta(n)\lvert n\in\{1,2,\cdots, N_T\}\right\},
\end{align}
where $\eta(n)\in S_{N}$. Here, $T_s$ denote the period of OFDM symbol. Fig.~2 sketches two different kinds of RSCS patterns: block-level and symbol-level. The first pattern shown in Fig.~2~(a) is called block-level pattern. Every block is made up of several even one thousand OFDM symbols. Within one block, the same RSCS pattern is used but the RSCS pattern will vary from one block to another. The second pattern shown in Fig.~2~(b) is a symbol-level type. In this type, the RSCS changes from one OFDM symbol to another. For desired receivers, the latter provides a more secure protection but also place a large computational complexity and uncertainty on the receive active subcarrier test. The former can strike a good balance among receiver complexity, security and performance by choosing a proper block size.  A large block size means a high successful detection probability of active subcarriers and less security because of less randomness. In other words, a better performance can be achieved. Determining the block size is a challenging problem beyond the scope of our paper due to the length limit on paper .
\begin{figure*}[htbp]
\centering
\subfigure[Block-level RSCS where RSCS pattern changes from one block to another with each block consisting of several OFDM symbols]{
\label{block}
\includegraphics[width=0.7\textwidth]{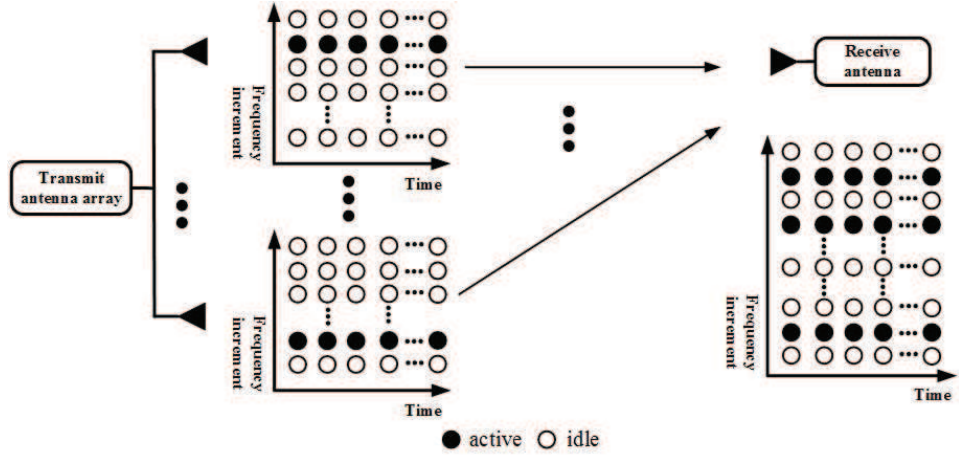}}
\hspace{1in}
\subfigure[Symbol-level RSCS where RSCS pattern varies from one OFDM symbol to another]{
\label{symbol}
\includegraphics[width=0.7\textwidth]{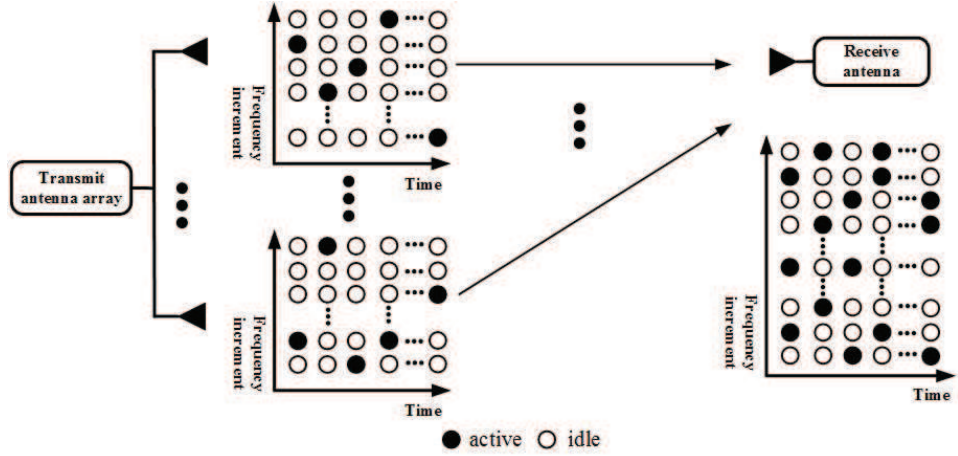}}
\caption{Concept of random-subcarrier-selection for transmitter and receiver.}
\label{fig2}
\end{figure*}
Let $\Delta T$ be the sampling interval, we have the next definition
\begin{align}
&T_s=N\Delta T=N/B,\\ \nonumber
&~~~~\Delta T=1/B.
\end{align}
With the above definitions, the transmit RF signal vector from $N_T$ transmit antennas is expressed in vector form as follows
\begin{align}
\mathbf{s}(t)=\left[s_1(t),s_2(t),\ldots,s_n(t),\ldots,s_{N_T}(t)\right]^T,
\end{align}
with
\begin{align}
s_n(t)=x_ke^{j(2\pi f_nt+\phi_n)},~(n=1,2,\ldots,N_T)
\end{align}
where $f_n$ is randomly chosen from subcarrier set $S_{sub}$, $x_k$ is the $k$th transmitted complex digital modulation symbol with $\mathbb{E}\{x_k^*x_k\}=1$, $\phi_n$ is the initial phase, and the time variable $t\in\left((k-1)T_s,kT_s\right)$. In far-field scenario, after the transmit signal $s_n(t)$ from element $n$ experiences the line-of -propagation (LoP) channel, the  corresponding receive signal at an arbitrary position  $(\theta,R)$, where $R$ and $\theta$ are the distance and  angle with respect to the first  element of transmit antenna array, respectively, and the first element is chosen as a reference antenna,  can be expressed as
\begin{align}
s'_n(\theta,R;t)&=x_ke^{j\left(2\pi f_n\left(t-\frac{R^{(n)}}{c}\right)+\phi_n\right)}\\ \nonumber
&=x_ke^{j\left(2\pi (f_c+\Delta f_{\eta(n)})\left(t-\frac{R^{(n)}}{c}\right)+\phi_n\right)},
\end{align}
with
\begin{align}
R^{(n)}=R-(n-1)d\cos\theta,
\end{align}
where $c$ is the light speed,  and $d=c/(2f_c)$ is the spacing between two elements of transmit antenna array, and $\Delta f_{\eta(n)}$ is the subcarrier frequency increment of the $n$th element. Hence, the overall received superimposed receive signal from all array elements  can be written as
\begin{align}\label{r_RF_signal}
&r_k'(\theta,R;t)=\rho\sum\limits_{n=1}^{N_T}s'_n(\theta,R;t)+n_k'(t)\\ \nonumber
&=\rho x_ke^{j2\pi f_c\left(t-\frac{R}{c}\right)}\\ \nonumber
&~~~~\cdot\sum\limits_{n=1}^{N_T}e^{j\left[2\pi \left(\Delta f_{\eta(n)} t-\Delta f_{\eta(n)}\frac{R^{(n)}}{c}+f_c\frac{(n-1)d\cos\theta}{c}\right)+\phi_n\right]}\\ \nonumber
&~~~~+n_k'(t),
\end{align}
where $\rho$ is the loss path factor in free space or LoP channel, and proportional to $\frac{1}{R^2}$, $n'_k(t)$ is the received channel noise. Therefore, the received useful signals at desired position $(\theta_D,R_D)$ and undesired position $(\theta_E,R_E)$ are expressed as
\begin{align}\label{y_u_d}
&r_k'(\theta_D,R_D;t)=\rho x_ke^{j2\pi f_c\left(t-\frac{R_D}{c}\right)}\\ \nonumber
&~~~~\cdot\sum\limits_{n=1}^{N_T}e^{j\left[2\pi \left(\Delta f_{\eta(n)}t-\Delta f_{\eta(n)}\frac{R_D^{(n)}}{c}+f_c\frac{(n-1)d\cos\theta_D}{c}\right)+\phi_n\right]}\\ \nonumber
&~~~~+n'_k(t),
\end{align}
and
\begin{align}\label{y_u_e}
&r_k'(\theta_E,R_E;t)=\rho x_ke^{j2\pi f_c\left(t-\frac{R_E}{c}\right)}\\ \nonumber
&~~~~\cdot\sum\limits_{n=1}^{N_T}e^{j\left[2\pi \left(\Delta f_{\eta(n)}t-\Delta f_{\eta(n)}\frac{R_E^{(n)}}{c}+f_c\frac{(n-1)d\cos\theta_E}{c}\right)+\phi_n\right]}\\ \nonumber
&~~~~+n'_k(t).
\end{align}
In what follows,  the received RF signal in (\ref{r_RF_signal}) is first down-converted to the following analog baseband signal
\begin{align}
&r_k(\theta,R,t)=\rho x_ke^{j2\pi f_c\left(t-\frac{R}{c}\right)}e^{-j2\pi f_c\left(t-\frac{R}{c}\right)}\\ \nonumber
&~~~~\cdot\sum\limits_{n=1}^{N_T}e^{j\left[2\pi \left(\Delta f_{\eta(n)}t-\Delta f_{\eta(n)}\frac{R^{(n)}}{c}+f_c\frac{(n-1)d\cos\theta}{c}\right)+\phi_n\right]}\\ \nonumber
&~~~~+n_k(t)\\ \nonumber
&=\rho x_k\sum\limits_{n=1}^{N_T}e^{j\left[2\pi \left(\Delta f_{\eta(n)}t-\Delta f_{\eta(n)}\frac{R^{(n)}}{c}+f_c\frac{(n-1)d\cos\theta}{c}\right)+\phi_n\right]}\\ \nonumber
&~~~~+n_k(t),
\end{align}
by using a signal down-conversion operation. Sampling the above analog complex baseband signal by signal bandwidth  $B$ Hz in complex field, then we get the received sampling signal sequence for $k$th OFDM symbol as follows
\begin{align}
\mathbf{r}_k[N]=\left[r_k[0],r_k[1],\ldots,r_k[m],\ldots,r_k[N-1]\right],
\end{align}
where $r_k[m]=r_k(t)|_{t=m\Delta T}$ denotes the $m$th sample of the signal with $\Delta T=\frac{1}{B}$ as follows
\begin{align}
&r_k[m]=\rho x_k\sum\limits_{n=1,\eta(n)\in S_{N}}^{N_T}\\ \nonumber
&\left(e^{j\left[2\pi \left(\Delta f_{\eta(n)}m\Delta T-\Delta f_{\eta(n)}\frac{R^{(n)}}{c}+f_c\frac{(n-1)d\cos\theta}{c}\right)+\phi_n\right]}\right)\\ \nonumber
&~~~~+n_k(m\Delta T)\\ \nonumber
&=\rho x_k\sum\limits_{n=1,\eta(n)\in S_{N}}^{N_T}\\ \nonumber
&\left(e^{j\left[2\pi\left(\eta(n)(\Delta fm\Delta T-\Delta f\frac{R-(n-1)d\cos\theta}{c})+\frac{(n-1)\cos\theta}{2}\right)+\phi_n\right]}\right)\\ \nonumber
&~~~~+n_k(m\Delta T),
\end{align}
where $\Delta f_{\eta(n)}=\eta(n)\Delta f$. Hence, the finite sequence $\mathbf{r}_k[N]$ has $N$ samples, and we divide the digital frequency interval $[0,~2\pi]$ into $N$ equidistant points. Thus, the form of $N$-points discrete Fourier transform (DFT) for sample signal is
\begin{align}
&y(q)=y(e^{j\omega})|_{\omega=2\pi q/N}=\sum\limits_{m=0}^{N-1}r_k[m]e^{-j\frac{2\pi}{N}mq}\\ \nonumber
&=\rho x_k\sum\limits_{m=0}^{N-1}\sum\limits_{n=1,\eta(n)\in S_{N}}^{N_T}e^{-j\frac{2\pi}{N}mq}\\ \nonumber
&\cdot e^{j2\pi\left(\eta(n)\Delta fm\Delta T-\eta(n)\Delta f\frac{R-(n-1)d\cos\theta}{c}+\frac{f_c(n-1)d\cos\theta}{c}\right)+j\phi_n}\\ \nonumber
&~~~~+\sum^{N-1}_{m=0}{n_k(m\Delta T)}e^{-j\frac{2\pi}{N}mq},
\end{align}
where $0\leq q\leq N-1$, and $\omega\in[0,~2\pi]$. Exchanging the order of double summations yields equation (\ref{head_y}).
\begin{figure*}
\begin{align}\label{head_y}
y(q)&=\sum\limits_{n=1,\eta(n)\in S_{N}}^{N_T}\left(\rho x_ke^{-j2\pi\left(\eta(n)\Delta f\frac{R-(n-1)d\cos\theta}{c}-\frac{f_c(n-1)d\cos\theta}{c}\right)+j\phi_n}\sum\limits_{m=0}^{N-1}e^{j2\pi\eta(n)\Delta fm\Delta T}e^{-j\frac{2\pi}{N}mq}+n_k(\eta(n)-q)\right)\\ \nonumber
&=\rho x_k\sum\limits_{n=1,\eta(n)\in S_{N}}^{N_T}e^{-j2\pi\left(\eta(n)\Delta f\frac{R-(n-1)d\cos\theta}{c}-\frac{f_c(n-1)d\cos\theta}{c}\right)+j\phi_n}\frac{\sin[\pi(\eta(n)-q)]}{\sin[\frac{\pi}{N}(\eta(n) -q)]}e^{j\frac{\pi}{N}(\eta(n)-q)(N-1)}+\sum^{N_T}_{n=1}{n_k(n)}.
\end{align}
\end{figure*}
Considering $0\le|(\eta(n)-q)|\le N$, we have
\begin{align}
&\frac{\sin[\pi(\eta(n)-q)]}{\sin[\frac{\pi}{N}(\eta(n)-q)]}e^{j\frac{\pi}{N}(\eta(n)-q)(N-1)}\\ \nonumber
&~~~~~~~~~~=\left\{
                                                                                               \begin{array}{ll}
                                                                                                 N,&\hbox{$q=\eta(n)$} \\
                                                                                                 0,&\hbox{$q\neq\eta(n)$}.
                                                                                               \end{array}
                                                                                             \right.
\end{align}
The received frequency-domain symbol over subchannel $q$ is rewritten as
\begin{align}\label{Rx_y_q_one}
&y(q)=\rho x_k\sum\limits_{n=1}^{N_T}\cdot\\ \nonumber
&\left(e^{-j2\pi\left(\Delta f_{\eta(n)}\frac{R^{(n)}}{c}-\frac{(n-1)\cos\theta}{2}\right)+j\phi_n}N\delta\left[2\pi\Delta f(\eta(n)-q)\right]\right)\\ \nonumber
&+\sum^{N_T}_{n=1}{n_k(n)},
\end{align}
with
\begin{align}
\delta(q=\eta(n))=\left\{
                                                                                               \begin{array}{ll}
                                                                                                 1,&\hbox{$q=\eta(n)$} \\
                                                                                                 0,&\hbox{$q\neq\eta(n)$}.
                                                                                               \end{array}
                                                                                             \right.
\end{align}
Based on (\ref{Rx_y_q_one}), all the received signals from $N_T$ subchannels are as follows
\begin{align}\label{Rx_sum_y_q}
&r_k(\theta,R,q)=\rho x_k\sum\limits_{n=1}^{N_T}\cdot\\ \nonumber
&\left(e^{-j2\pi\left(\eta(n)\Delta f\frac{R^{(n)}}{c}-\frac{(n-1)\cos\theta}{2}\right)+j\phi_n}N\delta\left[2\pi\Delta f(\eta(n)-q)\right]\right)\\ \nonumber
&+\sum^{N_T}_{n=1}{n_k(n)}\\ \nonumber
&=\mathbf{a}^H(\theta,R)\mathrm{diag}\{\mathbf{x}(q)\}\mathbf{v}_k+\sum^{N_T}_{n=1}{n_k(n)},
\end{align}
where the vector $\mathbf{a}^H(\theta,R)$ is as follows for a target at the angle $\theta$ and range $R$
\begin{align}\label{a}
&\mathbf{a}^H(\theta,R)\\ \nonumber
&~~=\left[e^{-j2\pi\psi_1},e^{-j2\pi\psi_2},...,e^{-j2\pi\psi_n},...,e^{-j2\pi\psi_{N_T}}\right],
\end{align}
where the function $\psi_n$ is defined by
\begin{align}
\psi_{n}=\eta(n)\Delta f\frac{R^{(n)}}{c}-f_c\frac{(n-1)d\cos\theta}{c}.
\end{align}
 In (\ref{Rx_sum_y_q}), the vector $\mathbf{v}_k$ is designed according to the knowledge of desired receiver position $(\theta_D,R_D)$ and defined as
\begin{align}
\mathbf{v}_k=\left[e^{j\phi_1},e^{j\phi_2},\ldots,e^{j\phi_{N_T}}\right]^T,
\end{align}
and
\begin{align}
&\mathbf{x}(q)=\rho x_kN\\ \nonumber
&~~~~\left\{\delta(2\pi\Delta f(\eta(1)-q)),\ldots,\delta(2\pi\Delta f(\eta(N_T)-q))\right\}.
\end{align}
Observing (\ref{Rx_sum_y_q}), to realize the constructive combining of  the $N_T$ terms at the desired receiver $(\theta_D, R_D)$, it is very apparent that all initial phases for $N_T$ transmit antennas satisfy the following identity
\begin{align}
&\phi_{1}-2\pi\psi_{D1}=\phi_{2}-2\pi\psi_{D2}=\cdots \phi_{N_T}-2\pi\psi_{DN_T}=\theta_0,
\end{align}
where $\theta_0$ is a constant phase, and
\begin{align}
\psi_{Dn}=\Delta f_{\eta(n)}\frac{R^{(n)}_D}{c}-f_c\frac{(n-1)d\cos\theta}{c}
\end{align}
with
\begin{align}
R^{(n)}_D=R_D-(n-1)d\cos\theta.
\end{align}
This condition will guarantee that all $N_T$ signals from $N_T$ antennas with random orthogonal frequency  shifts of the same signal is coherently combined at the desired receiver. The final receive combined signal at desired receiver is given by
\begin{align}\label{Rx_sum_r_d}
r_k(\theta_D,R_D)=\rho\zeta N_T x_k+\sum^{N_T}_{n=1}{n_k(n)},
\end{align}
where $\zeta=e^{j\theta_0}N$. Additionally, the corresponding signal at eavesdropper is
\begin{align}\label{Rx_sum_r_e}
&r_k(\theta_E,R_E)=\mathbf{a}^H(\theta_E,R_E)\mathrm{diag}\{\mathbf{x}(q)\}\mathbf{v}_k+\sum^{N_T}_{n=1}{n_k(n)}\\ \nonumber
&=\rho x_k\sum\limits_{n=1}^{N_T}e^{-j2\pi\left(\Delta f_{\eta(n)}\frac{R_E-(n-1)d\cos\theta}{c}-\frac{(n-1)\cos\theta_E}{2}\right)+j\phi_n}\\ \nonumber
&~~~~\cdot N\delta\left[2\pi\Delta f(\eta(n)-q)\right]+\sum^{N_T}_{n=1}{n_k(n)}.
\end{align}
Since all phases of signals received from all transmit antennas, $\phi_n-2\pi(\Delta f_{\eta(n)}\frac{R_E-(n-1)d\cos\theta_E}{c}-\frac{(n-1)\cos\theta_E}{2})$,  are viewed as independently identical distributed (iid) random variables due to random-subcarrier-selection,  the received signal sum $r_k(\theta_E,R_E)\rightarrow 0$ as $N_T$ tends to medium-scale or large-scale.

\section{Proposed AN-aided secure precise transmission method}
In the previous section, without the help of AN, we establish the principle of secure precise transmission by using RSCS, multiple transmit antennas, and phase alignment, which may transmit confidential messages to the given desired direction and distance with a weak signal leakage to other large region outside the small neighborhood around the given position. In this section, this scheme, in combination with AN, can further improve the security of transmitting confidential message. Below, we will show how to design the precoding vector for the confidential message and construct the AN projection matrix. Here, the AN projection matrix enforces the AN into the eavesdropper steering space or the null-space of the desired steering vector with a small fraction or no leakage of AN power  to the desired steering space. Conversely, the precoding vector for the confidential message is to drive the major power of confidential message to the desired steering space with a slight or no leakage to its null-space.
\subsection{General beamforming and AN projection scheme}
In terms of the derivation and analysis in Section II, we first construct the $k$ baseband frequency-domain vector corresponding to transmit antenna $m$ with other $N_T-1$ antennas being idle or non-active as follows
\begin{align}\label{Tx_S_km}
\mathbf{s}_{k,m}=\sqrt{P_S}\left[\beta_1v_{k}(m)x_k+\beta_2\tilde{w}(m)\right]\mathbf{e}_{\eta(m)},
\end{align}
where $P_S$ is the total transmit power, $\mathbf{e}_{\eta(n)}$ is an $N\times 1$ vector with only element $\eta(n)$ being one and others being zeros, $v_{k,m}$ denotes the $m$th element of the phase alignment/beamforming vector $\mathbf{v}_{k}$ which is to implement the coherent combining at the desired receiver, and $\tilde{w}_{m}$ denotes the $m$th element of the AN vector $\mathbf{\tilde{w}}$, which equals to $\mathbf{T}\mathbf{w}_k$. As we can see, $\mathbf{T}$ is an $N_T\times N_T$ orthogonal projection matrix and $\mathbf{w}_k\sim\mathcal{CN}(0,\mathbf{I}_N)$ is an $N_T\times 1$  random vector. In (\ref{Tx_S_km}), $\beta_1$ and  $\beta_2$ are two power allocation (PA)factors and satisfies the constraint $\beta_1^2+\beta_1^2=1$, which ensure that the total transmit power is equal to $P_S$. In accordance with (\ref{Tx_S_km}), the~$k$th total transmit spatial-frequency codeword from is rewritten as
\begin{align}\label{S}
\mathbf{S}_{k}=\sqrt{P_S}\beta_1\mathbf{E}_k\text{diag}\left(\mathbf{v}_{k}\right) x_k +\sqrt{P_S}\beta_2\mathbf{E}_k\text{diag}\left(\mathbf{T}\mathbf{w}_k\right),
\end{align}
where $\mathbf{E}_k$ is an $N\times N_T$ matrix of random-subcarrier-selection for transmit antenna array defined as
\begin{align}\label{E}
\mathbf{E}_k=\left[\mathbf{e}_{\eta(1)},~\cdots, \mathbf{e}_{\eta(m)},~\cdots,~\mathbf{e}_{\eta(N_T)} \right],
\end{align}
which is to force more AN into the null-space as possible as it can and at the same time less even no AN into the useful signal space consisting of the desired steering vectors. There are several various rules to optimize $\mathbf{v}_k$ and $\mathbf{T}$ like null-space projection, leakage and so on. Experiencing the LoP channel, the $N\times 1$ received vector of frequency-domain data symbols at position  $(\theta,R)$  is given by
\begin{align}
&\mathbf{y}_k(\theta,R)=\rho\sqrt{P_S}\sum_{m=1}^{N_T}\\ \nonumber
&~~a^*(\theta,R)(m)\left[\beta_1v_k(m)x_k+\beta_2\tilde{w}(m)\right]\mathbf{e}_{\eta(m)}+\mathbf{n}_k\\ \nonumber
&=\underbrace{\rho\sqrt{P_S}\beta_1\left\{\sum_{m=1}^{N_T}a^*(\theta,R)(m)\left[v_k(m)x_k\right]\mathbf{e}_{\eta(m)}\right\}x_k}_{\text{Useful~confidential~message}}\\ \nonumber
&+\underbrace{\rho\sqrt{P_S}\beta_2\sum_{m=1}^{N_T}a^*(\theta,R)(m)\tilde{w}(m)\mathbf{e}_{\eta(m)}}_{\text{Artifical~ noise}}+\underbrace{\mathbf{n}_k}_{\text{Channel~noise}}.
\end{align}
In the case of symbol-level precoding, making a summation of all $N$ subchannels yields
\begin{align}
&z_k(\theta,R)=\sum_{n=1}^{N}\mathbf{y}_k(\theta,R)(n)\\ \nonumber
&=\underbrace{\rho\sqrt{P_S}\beta_1\left\{\sum_{m=1}^{N_T}a^*(\theta,R)(m)v_k(m)\right\}x_k}_{\text{Useful~confidential~message}~ u_k(\theta,R)}\\ \nonumber
&+\underbrace{\rho\sqrt{P_S}\beta_2\sum_{m=1}^{N_T}a^*(\theta,R)(m)\mathbf{T}_m\mathbf{w}_k}_{\text{Artifical~noise}~\bar{w}_k(\theta,R)}+\underbrace{\sum_{n=1}^{N}\mathbf{n}_k(n)}_{\text{Channel~noise}~\bar{n}_k},
\end{align}
where the final projection vector $\mathbf{T}_{m}$ is the $m$th row of the AN projection matrix $\mathbf{T}$. The useful confidential signal $u_k(\theta,R)$, AN $\bar{w}_k(\theta,R)$, and channel noise $\bar{n}_k$ are assumed to be random variables with zero mean. Now, we define the average receive SINR as
\begin{align}\label{Average_SINR}
\text{SINR}(\theta,R)&=\frac{\mathbb{E}\{u_k(\theta,R)u^*_k(\theta,R)\}}{\mathbb{E}\{\bar{w}_k(\theta,R)\bar{w}^*_k(\theta,R)\}+\mathbb{E}\{\bar{n}_k\bar{n}_k^*\}}\\ \nonumber
&=\frac{\rho^2P_S\beta_1^2\|\mathbf{a}^H(\theta,R)\mathbf{v}_k\|^2}{\rho^2P_S\beta_2^2\|\mathbf{a}^H(\theta,R)\mathbf{T}\mathbf{w}_k\|^2+N\sigma^2_n},
\end{align}
which is also proved to be valid for $\mathbb{E}\{\bar{w}_k(\theta,R)\bar{w}^*_k(\theta,R)\}\approx$ $\mathbb{E}\{\bar{n}_k\bar{n}_k^*\}$ in Appendix A.

In the block-level precoding case, assume the receiver at position $(\theta,R)$  is so smart that it can identify the set of the $N_T$ active subcarrier indices by energy detection algorithm similar to cognitive radio \cite{harry}, the corresponding  $N_T$ active subchannels are summed together
\begin{align}
&z_k(\theta,R)\\ \nonumber
&~~~~=\underbrace{\rho\sqrt{P_S}\beta_1\left\{\sum_{m=1}^{N_T}a^*(\theta,R)(m)v_k(m)\right\}x_k}_{\text{Useful~confidential~message}}\\ \nonumber
&~~~~~+\underbrace{\rho\sqrt{P_S}\beta_2\sum_{m=1}^{N_T}a^*(\theta,R)(m)\mathbf{T}_m\mathbf{w}_k}_{\text{Artifical~ noise}}+\underbrace{\sum_{n=1}^{N_T}\mathbf{n}_k(n)}_{\text{Channel~noise}}.
\end{align}
According to the above model and similar to Appendix A, the average SINR is given by
\begin{align}\label{Smart_Exact_SINR}
\text{SINR}(\theta,R)=\frac{\rho^2P_S\beta_1^2\|\mathbf{a}^H(\theta,R)\mathbf{v}_k\|^2}{\rho^2P_S\beta_2^2\|\mathbf{a}^H(\theta,R)\mathbf{T}\mathbf{w}_k\|^2+N_T\sigma^2_n}.
\end{align}

\subsection{Analysis of receive SINR performance in the case of phase alignment and null-space projection}
In particular, when the ideal desired direction angle and distance are available, we use a simple beamforming form
\begin{align}\label{PA_Precoding}
\mathbf{v}_k=\frac{1}{\sqrt{N_T}}\mathbf{a}(\theta_D,R_D),
\end{align}
which provides a maximum coherent combining at desired receiver, and
\begin{align}
\mathbf{a}^H(\theta_D,R_D)\mathbf{T}=\mathbf{0}_{1\times N_T},
\end{align}
which projects the AN onto the null-space of the desired steering vector $\mathbf{a}(\theta_D,R_D)$. For simplification, $\mathbf{T}_m$ is taken to be the $m$th row of the following projection matrix
\begin{align}\label{AN_NSP}
\mathbf{T}=\mathbf{I}_{N_T}-\frac{1}{N_T}\mathbf{a}(\theta_D,R_D)\mathbf{a}^H(\theta_D,R_D).
\end{align}
By using the above beamforming scheme, the desired receive coherent combining signal is
\begin{align}
z_k(\theta_D,R_D)=\underbrace{\rho\sqrt{P_SN_T}\beta_1x_k}_{\text{Useful~confidential~message}}+\underbrace{\sum_{n=1}^{N_T}\mathbf{n}_k(n)}_{\text{Channel~noise}},
\end{align}
Due to phase alignment at transmitter, that is, constant-envelop beamformer with all elements having the same magnitude, we have the following average SINR
\begin{align}\label{Exact_SINR_d}
\text{SINR}(\theta_D,R_D)=\frac{\rho^2P_S\beta_1^2N_T}{N_T\sigma^2_n}=\frac{\rho^2P_S\beta_1^2}{\sigma^2_n},
\end{align}
since the AN projection matrix at desired receive position is perpendicular to the desired steering vector. At the eavesdroppers, the receive combining signal is
\begin{align}
&z_k(\theta_E,R_E)=\underbrace{\rho\sqrt{\frac{P_S}{N_T}}\beta_1\mathbf{a}^H(\theta_E,R_E)\mathbf{a}(\theta_D,R_D)x_k}_{\text{Useful~confidential~message}}\\ \nonumber
&+\underbrace{\rho\sqrt{P_S}\beta_2\mathbf{a}^H(\theta_E,R_E)\mathbf{T}\mathbf{w}_k}_{\text{Artifical~ noise}}+\underbrace{\sum_{n=1}^{N_T}\mathbf{n}_k(n)}_{\text{Channel~noise}}.
\end{align}
The SINR expression at undesired position $(\theta_E,R_E)$ reduces to
\begin{align}\label{Exact_SINR_e}
&\text{SINR}(\theta_E,R_E)\\ \nonumber
&~~~~=\frac{\rho^2P_S\beta_1^2\|\mathbf{a}^H(\theta_E,R_E)\mathbf{v}_k\|^2}{\rho^2P_S\beta_2^2\|\mathbf{a}^H(\theta_E,R_E)\mathbf{T}\mathbf{w}_k\|^2+N_T\sigma^2_n}.
\end{align}
In a practical system, it is preferred that the SINR at desired position is maximized given that the SINR at eavesdropper position $\text{SINR}(\theta_E,R_E)$ is lower than the predefined value like 0dB.

\subsection{Analysis of average secrecy rate}
With the help of the definition of average SINR in the previous subsection, the average secrecy rate of the system is written as
\begin{align}\label{Average-SR}
&SR=\log_2(1+\text{SINR}(\theta_D,R_D))\\ \nonumber
&~~~~-\max\limits_{(\theta_E,R_E)\in \text{wiretap~area}}\{\log_2(1+\text{SINR}(\theta_E,R_E))\} \\ \nonumber
&\text{subject~to}:~~\text{wiretap~area}=\\ \nonumber
&~~~~\left\{(\theta_E,R_E)\big{|}|R_E-R_D|\geq \Delta R,~|\theta_E-\theta_D|\geq\Delta\theta\right\},
\end{align}
where $\Delta R$ and $\Delta\theta$ can refer to the first null position around the main beam of the desired receiver, which is defined in Appendix B. The constraint in the above definition denotes the wiretap region where its complementary set is the desired region. The above definition of average secrecy rate means the achievable rate at the desired receiver minus the maximum rate achievable by eavesdroppers within the wiretap region. In Appendix B, we derive a tight upper bound on (\ref{Exact_SINR_e})
\begin{align}
&\text{SINR}(\theta_E,R_E)\\ \nonumber
&~~=\frac{\rho^2P_S\beta_1^2\|\mathbf{a}^H(\theta_E,R_E)\mathbf{v}_k\|^2}{\rho^2P_S\beta_2^2\|\mathbf{a}^H(\theta_E,R_E)\mathbf{T}\mathbf{w}_k\|^2+N_T\sigma_n^2}\\ \nonumber
&~~=\frac{\rho^2P_S\beta_1^2\|\mathbf{a}^H(\theta_E,R_E)\mathbf{a}(\theta_D,R_D)\|^2}{N_T\rho^2P_S\beta_2^2\|\mathbf{a}^H(\theta_E,R_E)\mathbf{T}\mathbf{w}_k\|^2+N_T^2\sigma_n^2}.
\end{align}
as follows
\begin{align}\label{EaveSINR_UB}
&\text{SINR}(\theta_E,R_E)\\ \nonumber
&~~~~\leq\max\left\{\text{SINR}(\lambda_{\max1}),~\text{SINR}(\lambda_{\max2})\right\},
\end{align}
which is achievable and a tight upper bound, where parameters $\lambda_{\max1}$ and  $\lambda_{\max2}$ are given by
\begin{align}
\lambda_{\max1}=&\frac{1}{N_T}\sum_{n=1}^{N_T}e^{-j2\pi(\Delta f_{\eta(n)}\frac{R_{E_{\mathrm{sidelobe}}}-R_D}{c})}\\ \nonumber
&\cdot\frac{1}{N_T}\sum_{n=1}^{N_T}e^{j2\pi(\Delta f_{\eta(n)}\frac{R_{E_{\mathrm{sidelobe}}}-R_D}{c})},\\
\lambda_{\max2}=&\frac{1}{N_T}\sum_{n=1}^{N_T}e^{j2\pi(\frac{1}{2}\left(\cos\theta_{E_{\mathrm{sidelobe}}}-\cos\theta_D\right)(n-1))}\\ \nonumber
&\cdot\frac{1}{N_T}\sum_{n=1}^{N_T}e^{-j2\pi(\frac{1}{2}\left(\cos\theta_{E_{\mathrm{sidelobe}}}-\cos\theta_D\right)(n-1))},
\end{align}
where $\theta_{E_{\mathrm{sidelobe}}}$ and $R_{E_{\mathrm{sidelobe}}}$are defined in Appendix B. Its detailed proving process is presented in Appendix B. Evidently, by adjusting the values of parameters $\beta_1$ and $\beta_2$, we may make the SINR value low enough to be unable to be correctly detected by eavesdropper.

\textbf{Theorem~1:} In a secure precise communication shown in Fig.~1, if the phase-alignment precoder for confidential messages in (\ref{PA_Precoding})  is adopted and the AN null-space projection method in (\ref{AN_NSP}) is used, the average secrecy rate  is approximated by the following formula
\begin{align}\label{Average-SR-Final}
&SR=\log_2\left(1+\text{SINR}(\theta_D,R_D)\right)\\ \nonumber
&~~~~-\log_2\left(1+\max\left\{\text{SINR}(\lambda_{\max1}),~\text{SINR}(\lambda_{\max2})\right\}\right),
\end{align}
where the first term in the right hand side of the above equation denotes the rate achieved by desired user and the second one is the  upper bound of rate achievable by eavesdropper.
\hfill$\blacksquare$

{\emph{Proof}: The proof of Theorem~1 is straightforward by combining (\ref{Average-SR}), and (\ref{EaveSINR_UB})}.\hfill$\blacksquare$

\section{Simulation results and analysis}
In what follows, the proposed scheme RSCS-OFDM-DM is evaluated in terms of the average  SINR and secrecy rate performance. In our simulation, system parameters are chosen as follows: the carrier frequency $f_c=3$GHz, the total signal bandwidth $B$ changes from $5$MHz to $100$MHz, the number of total subcarriers $N=1024$, $P_S/\sigma_n^2=10$dB,  the antenna element spacing is half of the wavelength (i.e., $d=c/2f_c$), the position of the desired receiver is chosen to be ($60^\circ,500m$).

\begin{figure}[h]
\centering
\includegraphics[width=9cm]{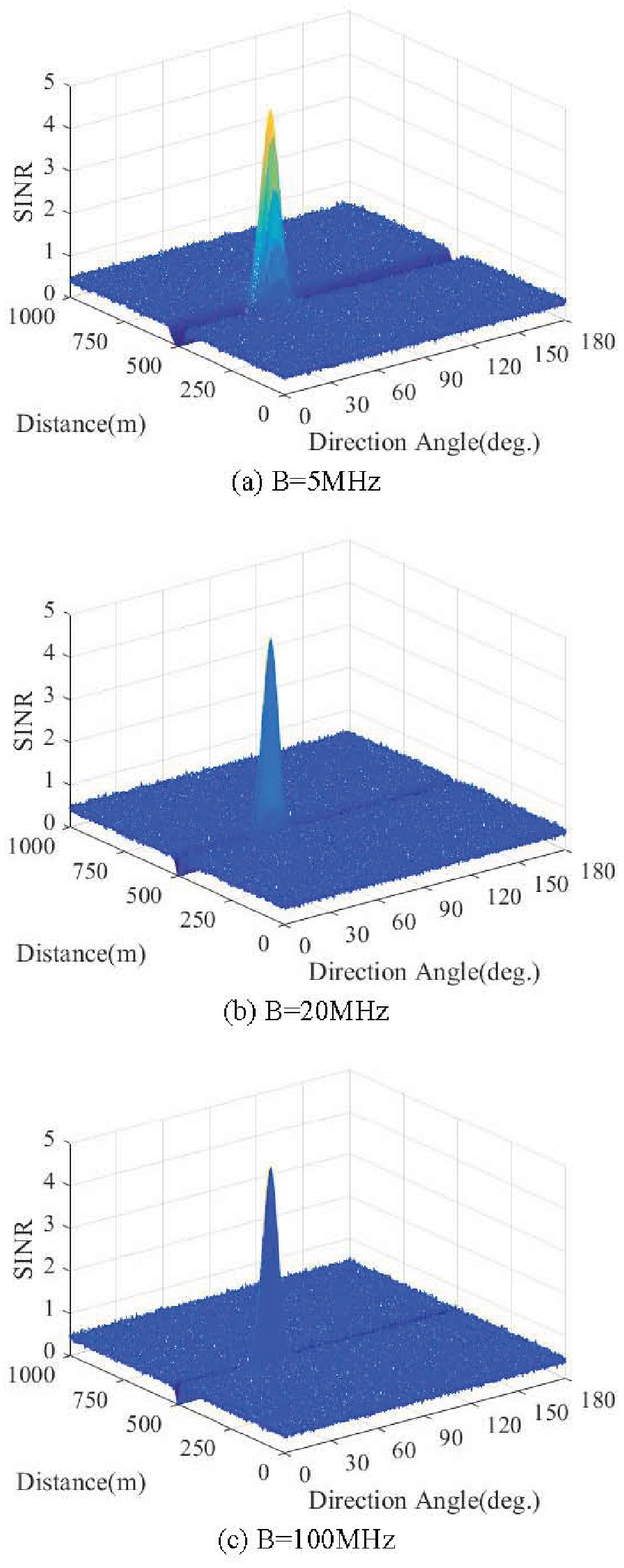}
\caption{3-D surface of SINR versus direction angle and distance of the proposed method for $N_T=8$ ($P_S/\sigma_n^2=10$dB,~$\beta_1^2=\beta_2^2=0.5$).}
\label{fig3}
\end{figure}

Fig.~\ref{fig3} illustrates the $3$-D performance surface of SINR versus direction angle $\theta$ and distance $R$ of the proposed  method of phase alignment/beamforming plus null-space projection in Section III for three different bandwidths: 5MHz, 20MHz, and 100MHz. Observing three parts in Fig.~3,  the receive power forms a high useful signal energy peak only around the desired receiver position $(60^\circ,500m)$ due to the joint use of AN projection and PA. Otherwise, in other undesired region of excluding the main useful power peak, the receive SINR is so low that it is smaller than one, due to a weak receive confidential signal corrupted by AN.  Additionally, we also find an very interesting fact that the width of useful main peak along distance dimension decreases accordingly as the signal bandwidth grows from 5MHz to 100MHz.  In summary, a larger signal bandwidth  generates a narrower confidential power peak along distance dimension but has no influence on the width of peak along direction dimension. This means that a larger security along distance dimension can be offered by using a larger signal bandwidth.

\begin{figure}[h]
\centering
\includegraphics[width=9cm]{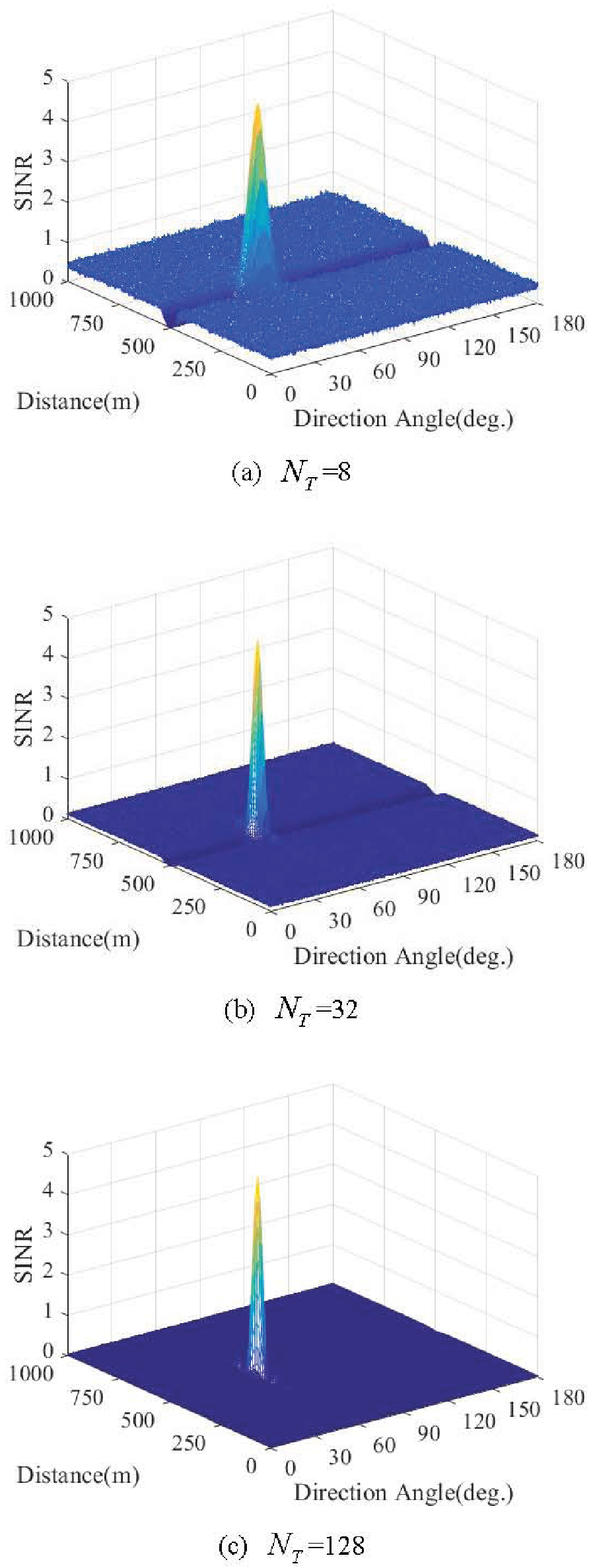}
\caption{3-D surface  of SINR versus direction angle and distance of the proposed method for $B$=5MHz ($P_S/\sigma_n^2=10$dB,~$\beta_1^2=\beta_2^2=0.5$).}
\label{fig4}
\end{figure}

Fig.~\ref{fig4} draws the $3$-D performance surface  of SINR versus direction angle and distance of the proposed method for three different numbers of transmit antennas: 8, 32 and 128. Via three parts (a), (b), and (c) in  Fig.~4, we find the following fact that only single energy peak  is synthesized around the desired receiver position $(60^\circ,500m)$ similar to Fig.~\ref{fig3} by beamforming and AN projection. Furthermore, with the increasing in the number of transmit antennas, the width of the peak along direction dimension becomes narrower and narrower. This can be readily explained. Given a fixed antenna spacing, increasing the number of transmit antenna array will increase antenna array size and provide a high-spatial-angle-resolution, i.e., a narrower peak along direction dimension.  As the number of transmit antennas tends to large-scale, the peak converges to an extremely narrow peak along direction dimension and perpendicular to the plane along direction dimension. If both  large signal bandwidth $B$ and large-scale transmit antenna array are used, the  peak is formed to narrow  along both direction and distance dimensions. Then, the ultimate aim of the true secure precise transmission is achieved in terms of the outcomes in Fig.~3 and Fig.~4. That is, the peak converges to a line perpendicular to the 2-D plane of direction and distance dimensions and at the desired position.

\begin{figure}[h]
\centering
\includegraphics[width=9cm]{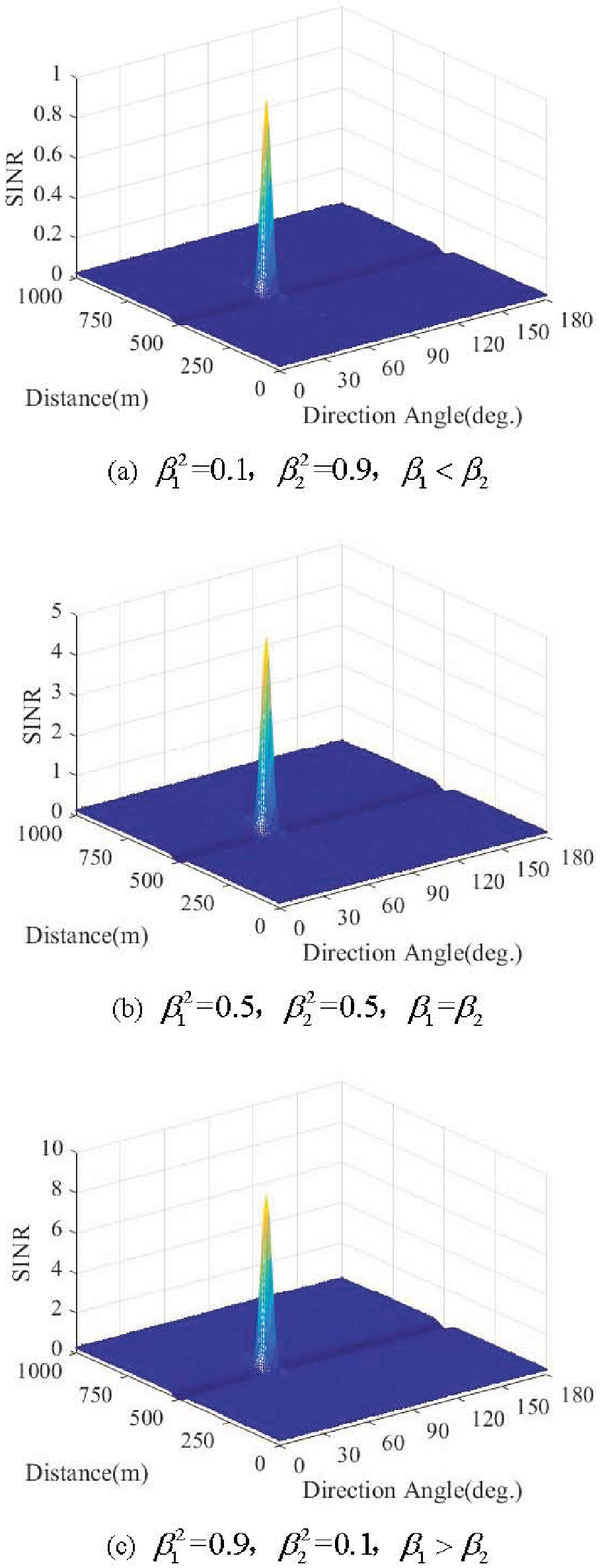}
\caption{3-D surface of SINR versus direction angle and distance of the proposed method for $B$=5MHz, and $N_T=32$ ($P_S/\sigma_n^2=10$dB).}
\label{fig5}
\end{figure}

\begin{figure}[htp]
\centering
\subfigure[$N_T=8$]{
\includegraphics[width=0.45\textwidth]{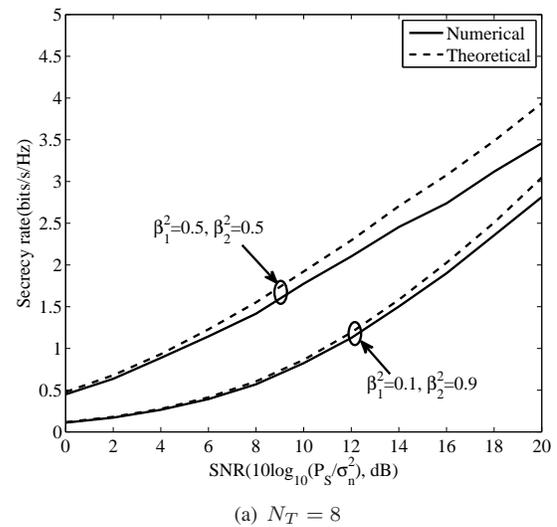}}
\hspace{1in}
\subfigure[$N_T=32$]{
\includegraphics[width=0.45\textwidth]{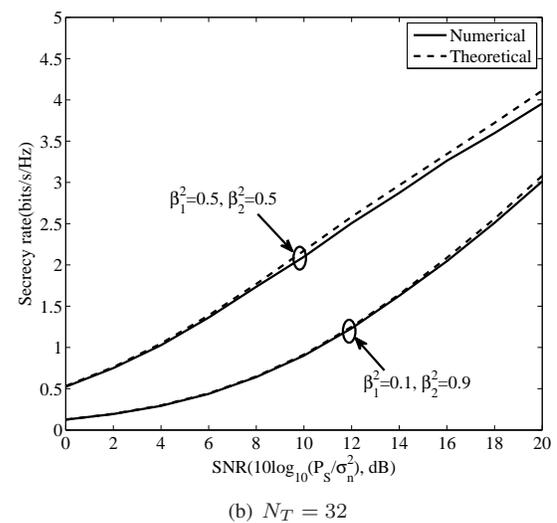}}
\hspace{1in}
\subfigure[$N_T=128$]{
\includegraphics[width=0.45\textwidth]{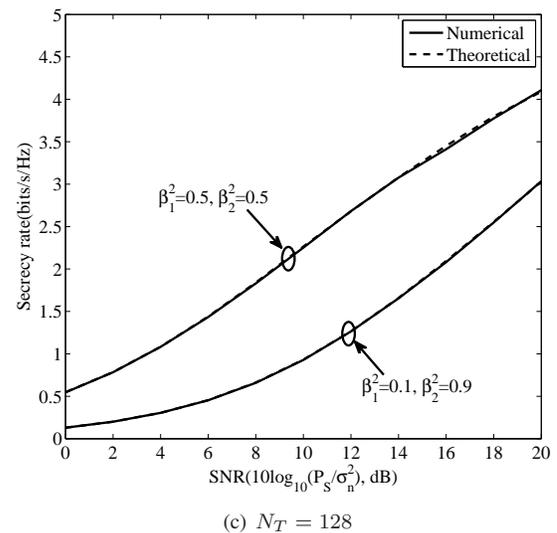}}
\caption{Curves of secrecy rate  versus SNR($10\log_{10}(P_S/\sigma_n^2)$) of the proposed scheme, where $B=5$MHz.}\label{fig6}
\end{figure}

\begin{figure}[htp]
\centering
\subfigure[SNR ($10\log_{10}(P_S/\sigma_n^2)$) = 0dB]{
\includegraphics[width=0.45\textwidth]{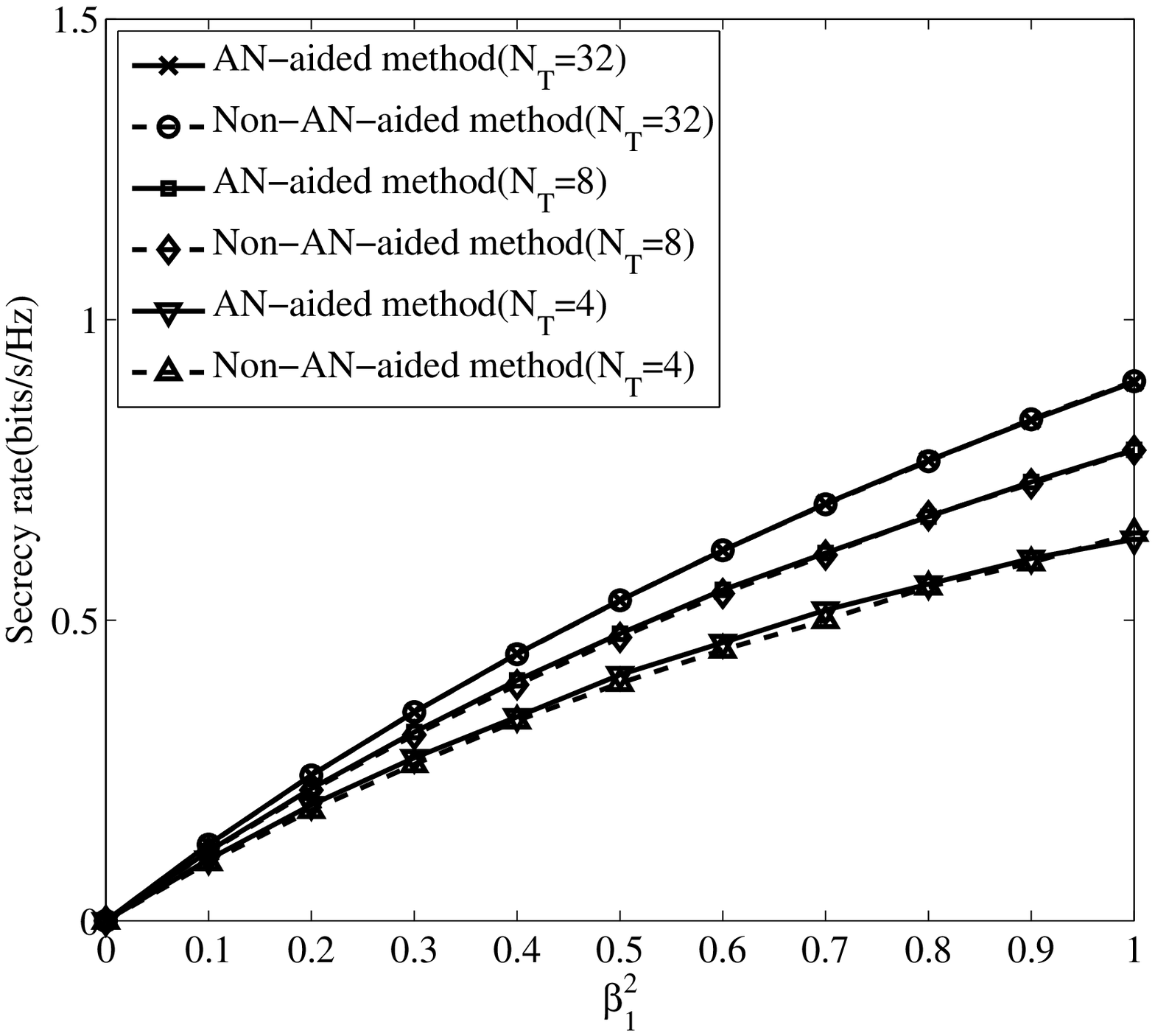}}
\hspace{1in}
\subfigure[SNR ($10\log_{10}(P_S/\sigma_n^2)$) = 10dB]{
\includegraphics[width=0.45\textwidth]{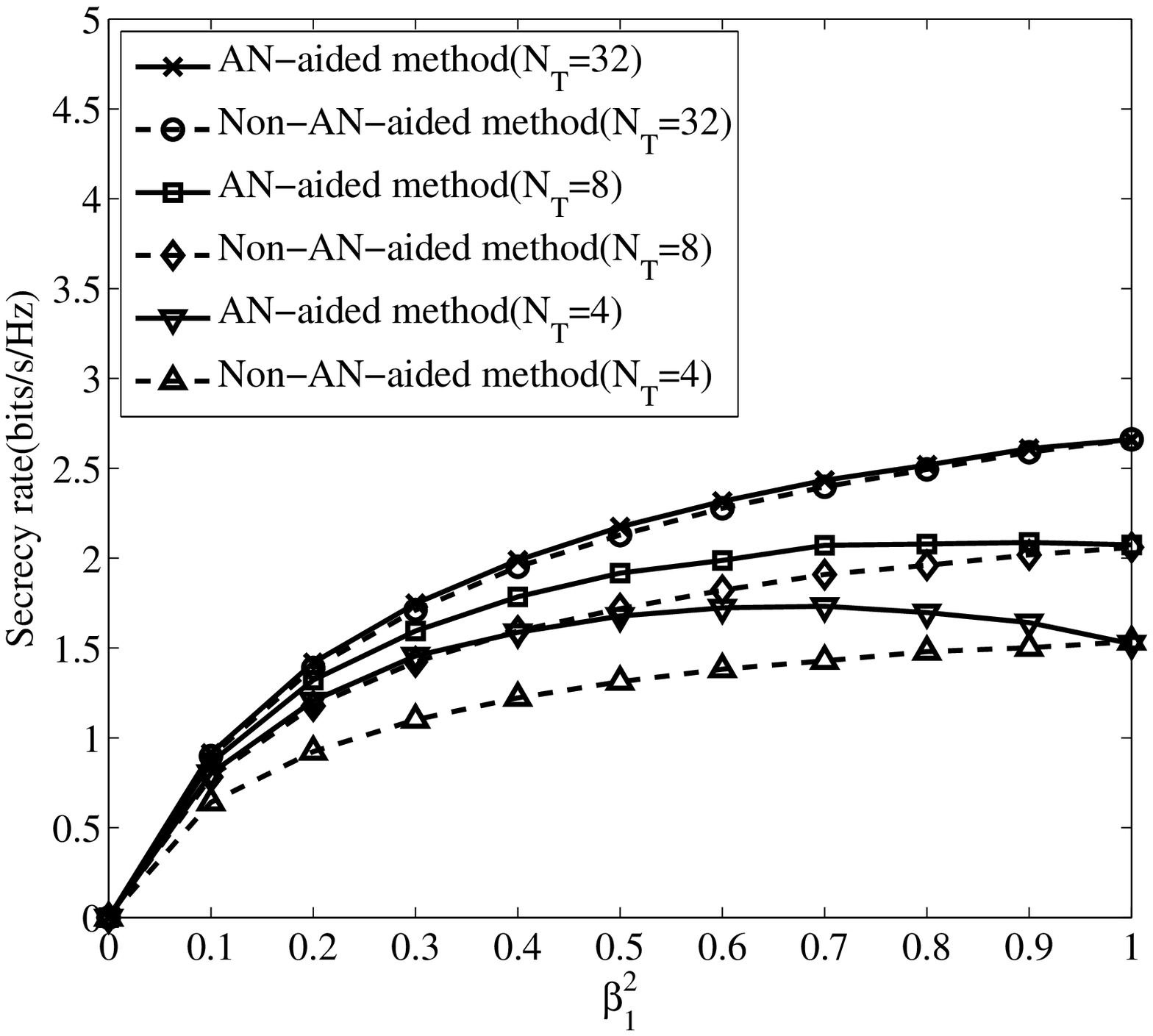}}
\hspace{1in}
\subfigure[SNR ($10\log_{10}(P_S/\sigma_n^2)$) = 20dB]{
\includegraphics[width=0.45\textwidth]{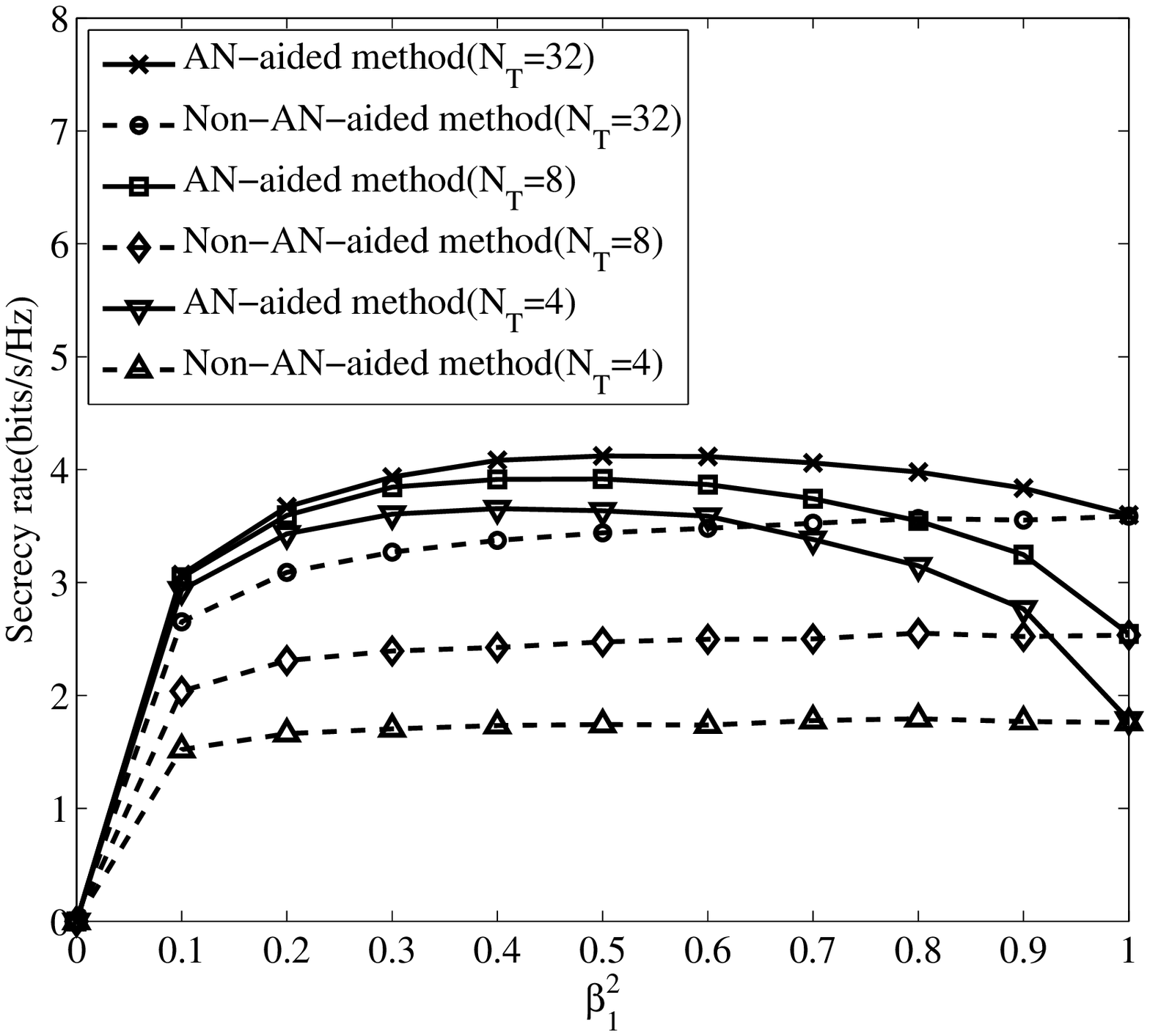}}
\caption{Curves of theoretical and numerical average secrecy rate versus $\beta_1^2$, where $B=5$MHz.}
\label{fig7}
\end{figure}

In Fig.~5, we will evaluate the impact of power allocation parameters $\beta_1$ and $\beta_2$ on the SINR performance. Three different typical  scenarios are considered as follows: (1)~$\beta_1<\beta_2$ with $\beta_1^2=0.1$ and $\beta_2^2=0.9$, (2)~$\beta_1=\beta_2$, with $\beta_1^2=\beta_2^2=0.5$, and (3)~$\beta_1>\beta_2$ with $\beta_1^2=0.9$ and $\beta_2^2=0.1$, which corresponds to three different parts in Fig.~5. From the three parts, the maximum value of SINR peak decreases as the value of $\beta_1$ decreases. The result is obvious intuitively. Also, the SINR value outside the main peak approaches zero. Thus, increasing the value of parameter $\beta_2$ will further degrade the SINR performance of eavesdropper regions. Conversely, increasing the value of parameter $\beta_1$ will improve the SINR performance of desired receiver but also lower the security. Additionally, the values of $\beta_1$ and $\beta_1$ are also intimately related to energy efficiency. Thus, how to choose the values of $\beta_1$ and $\beta_1$ depends on application scenarios and should consider a balance among energy efficiency, security and performance.

  Fig.~6 demonstrates the curves of derived theoretical and numerical average secrecy rates versus SNR for three different numbers $N_T$ of transmit antennas.  Parts (a), (b), and (c) correspond to $N_T=8,~32, 128$, respectively, where each part includes two different power allocation strategies. In this figure, the derived theoretical curve is given by (\ref{Average-SR}) while the simulated numerical curve is the exact SR values by sufficient Monte Carlo simulation runs. Observing the three parts, the gap of the simulated curve and the theoretical one disappears as the number of transmit antennas increases from $8$ to $128$. Thus, the derived theoretical SR expression achieves a good approximation to the exact SR. This also verifies the validness of our expression in (\ref{Average-SR}).  Additionally,  in the low and medium SNR regions, the theoretical curve is also very close to  the numerical one such that they overlap together for three parts.


Below, we make an investigation concerning the impact of power allocation parameters $\beta_1$ and $\beta_2$  on the secrecy rate performance in Fig.~7. Parts (a), (b), and (c) illustrate the secrecy rate versus $\beta_1$ in three different SNR scenarios: (1) SNR=0dB, (2) SNR=10dB, and (3) SNR=20dB, respectively. Observing the three parts, we first find that the gap between AN-aided method and non-AN-aided method decreases as the number of transmit antennas increases, where the non-AN-aided implies $\beta_2=0$ and no AN. In Part (a), that is, in low SNR region and with large number of antennas, the influence of AN is weak due to a large channel noise. However, as SNR increases , the AN becomes more and more important.  In Parts (b) and (c),  as the value of $\beta_1$ varies from 0 to 1, the average SR curve  is a concave function of $\beta_1$. There exists an optimal power allocation strategy of maximizing the SR.  As SNR increases, the optimal power allocation factor $\beta_1$ decreases. A high SNR means a good desired channel quality. In other words, for a high SNR scenario,  more power may be used to transmit artificial noise(AN) to corrupt the eavesdroppers while less power is adopted to transmit confidential message.

\section{Conclusion}
In this paper, the RSCS-OFDM-DM scheme is proposed to achieve an ultimate goal of secure precise transmission of confidential messages. To accomplish this goal, several important tools are utilized like: random subcarrier selection, phase alignment/beamforming for confidential messages, and null-space projection of AN. Using this scheme, we obtain the following interesting results: 1) the proposed scheme can generate a high SINR peak only at the desired position $(\theta_D,R_D)$ and a low flat SINR plane for other eavesdropper regions with SINR being far less than the former, 2) the widths of main peak along angle and distance dimensions are intimately related to  the number of antennas $N_T$ and  bandwidth $B$, respectively. Increasing $B$ and $N_T$ will achieve a narrower and narrower main peak along both direction and distance dimensions at desired position, 3) increasing the fraction of AN power in the total transmit power will gradually degrade the SINR performance of eavesdropper at the expense of that of desired user. The final result is as follows: at eavesdroppers, the received power of confidential messages is very weak and corrupted by AN, thus their receive SINR is so poor that the confidential messages can not be successfully and reliably intercepted outside main peak at the desired position, and the transmit confidential power mainly gathers in a small neighbourhood around the desired position, where only a small fraction of the total power leaks out to eavesdropper regions. The proposed RSCS-OFDM-DM scheme is easy to be implemented compared to the conventional RFDA proposed in \cite{Hu2}. Finally, we also derive the tight upper bound for SINR in wiretap region. By using this bound, we attain the approximate expression for  average secrecy rate. Simulation results and analysis confirm that the derived theoretical expression of average secrecy rate is a good approximation to the exact average SR. Due to the low-complexity and high security of the proposed scheme, it will be potentially applied in the future scenarios including unmanned aerial vehicle communications, satellite communications, mm-Wave communications and so on.

\appendices
\section{Approximate Derivation of Mean of SINR$(\theta,R)$ for $\sigma^2_{CN}/2\approx\sigma^2_{AN}/2$}
In the position $(\theta,R)$, consider
\begin{align}\label{u_k_expression}
u_k(\theta,R)=\rho\sqrt{P_S}\beta_1\left\{\sum_{m=1}^{N_T}a^*(\theta,R)(m)v_k(m)\right\}x_k,
\end{align}
and
\begin{align}\label{g_k_expression}
&g_k(\theta,R)=\bar{w}_k(\theta,R)+\bar{n}_k\\ \nonumber
&=\underbrace{\rho\sqrt{P_S}\beta_2\sum_{m=1}^{N_T}a^*(\theta,R)(m)\mathbf{T}_m\mathbf{w}_k}_{g_{k,AN}(\theta,R)}+\underbrace{\sum_{n=1}^{N_T}\mathbf{n}_k(n)}_{g_{k,CN}(\theta,R)}.
\end{align}
where channel noise is abbreviated as CN. Firstly, let us address the problem of PDF of random variable $u_k(\theta,R)$. Three random variables $u_k(\theta,R)$, $g_{k,AN}(\theta,R)$, and $g_{k,CN}(\theta,R)$ are the sum of $N_T$ random variables. In order to simplify the following deriving, they are approximately modelled as complex Gaussian distributions in terms of the central limit theorem in probability theory. From  (\ref{u_k_expression}) and (\ref{g_k_expression}), their means and variances are
\begin{align}\label{ug_k_expectation}
&\mathbb{E}\{u_k(\theta,R)\}=0,~\mathbb{E}\{g_{k,AN}(\theta,R)\}=0,\\ \nonumber
&~~~~~~~~\mathbb{E}\{g_{k,CN}(\theta,R)\}=0,
\end{align}
and
\begin{align}\label{u_k_variance}
\sigma_u^2&=\mathbb{E}\{u_k(\theta,R)u_k^*(\theta,R)\}=\rho^2P_S\beta_1^2\|\mathbf{a}^H(\theta,R)\mathbf{v}_k\|^2,
\end{align}
\begin{align}\label{g_k_variance}
&\sigma^2_{AN}=N_TP_S\rho^2\beta_2^2\sigma^2_{w}\sum_{m=1}^{N_T}\|a^*(\theta,R)(m)\|^2\mathbf{T}_m\mathbf{T}^H_m,\\ \nonumber
&\sigma_{CN}^2=N_T\sigma^2_n.
\end{align}
Considering the real-time SINR $\gamma$ is approximately given by
\begin{align}\label{Realtime_SINR}
\gamma=\frac{u_k(\theta,R)u^*_k(\theta,R)}{g_{k,AN}(\theta,R)g_{k,AN}^*(\theta,R)+g_{k,CN}g_{k,CN}^*}.
\end{align}
In accordance with the above definition, let define the three new random variables as follows
\begin{align}
&\alpha=\frac{u_k(\theta,R)u_k^*(\theta,R)}{\sigma_u^2/2},\\ \nonumber
&\beta=\frac{g_{k,AN}(\theta,R)g_{k,AN}^*(\theta,R)}{\sigma_{AN}^2/2},\\ \nonumber
&\lambda=\frac{g_{k,CN}g_{k,CN}^*}{\sigma_{CN}^2/2},
\end{align}
which are three Chi-squared distribution with two degrees of freedom. Without losing generality, putting the above variables in (\ref{Realtime_SINR}) yields
\begin{align}\label{gamma}
\gamma=\frac{\frac{\sigma_u^2}{2}\alpha}{\frac{\sigma_{AN}^2}{2}\beta+\frac{\sigma_{CN}^2}{2}\lambda}=e\bullet\frac{\alpha}{a\beta+b\lambda},
\end{align}
where $e=\sigma^2_{u}/2$, $a=\sigma^2_{AN}/2$, $b=\sigma^2_{CN}/2$, three random variables $\alpha$, $\beta$ and $\lambda$ have the same PDF as follows \cite{kay}
\begin{align}
f_{\alpha}(x)=f_{\beta}(x)=f_{\lambda}(x)=\frac{1}{2}\exp(-\frac{1}{2}x).
\end{align}
Below, we discuss one typical situation: ~$\sigma^2_{CN}/2\approx\sigma^2_{AN}/2$. In this scenario,~we will give an approximation of the PDF of random variable $\gamma$. The distribution of $\frac{a\beta+b\lambda}{(a+b)}$ is approximately regarded as one Chi-squared distribution with four degrees of freedom. Then, the SINR in (\ref{gamma}) is rewritten by
\begin{align}\label{approximately_gamma}
\gamma=\frac{\frac{\sigma^2_{u}}{2}\alpha}{\frac{\sigma^2_{AN}}{2}\beta+\frac{\sigma^2_{CN}}{2}\lambda}=\frac{e\alpha}{a\beta+b\lambda}=\frac{e}{2(a+b)}\bullet\underbrace{\frac{\alpha/2}{q/4}}_{\tilde{\gamma}},
\end{align}
From the definition of $F$-distribution, we have the PDF of $\tilde{\gamma}$ as follows
\begin{align}\label{tilde_gamma}
f_{\tilde{\gamma}}(x)=\frac{1}{\left(1+\frac{1}{2}x\right)^3},
\end{align}
which leads to the PDF of random variable $\gamma$
\begin{align}\label{f_gamma}
f_{\gamma}(x)=\frac{2(a+b)}{e}\frac{1}{\left(1+\frac{(a+b)}{e}x\right)^3},~x>0
\end{align}
which yields the average value of SINR $\gamma$
\begin{align}\label{mean_gamma}
\bar{\gamma}=\mathbb{E}\{\gamma\}=\frac{2(a+b)}{e}\int^{+\infty}_{0}\frac{x}{\left(1+\frac{(a+b)}{e}x\right)^3}dx.
\end{align}
Let us define a new integral variable $x'=\left(1+\frac{(a+b)}{e}x\right)$, then the above integral can be converted into
\begin{align}\label{approximately_mean}
\bar{\gamma}=\frac{2e}{a+b}\int^{+\infty}_{1}\frac{x'-1}{\left(x'\right)^3}dx'=\frac{e}{a+b}.
\end{align}
by checking the integral table in \cite{grad,wasser}. This completes the proof of the PDF of SINR and its mean.
\hfill$\blacksquare$\label{Append_A}

\section{Tight upper bound of $\text{SINR}(\theta_E,R_E)$}

In this Appendix, a tight upper bound for the receive SINR in eavesdropper region is proved. By means of the SINR definition in (\ref{Exact_SINR_e}),  the SINR $\text{SINR}(\theta_E,R_E)$ is rewritten as
\begin{align}
&\text{SINR}(\theta_E,R_E)=\frac{\rho^2P_S\beta_1^2\|\mathbf{a}^H(\theta_E,R_E)\frac{1}{\sqrt{N_T}}\mathbf{a}(\theta_D,R_D)\|^2}{\rho^2P_S\beta_2^2\|\mathbf{a}^H(\theta_E,R_E)\mathbf{T}\mathbf{w}_k\|^2+N_T\sigma_n^2}\\ \nonumber
&~~~~=\frac{\rho^2P_S\beta_1^2/N_T\|\mathbf{a}^H(\theta_E,R_E)\mathbf{a}(\theta_D,R_D)\|^2}{\rho^2P_S\beta_2^2\|\mathbf{a}^H(\theta_E,R_E)\mathbf{T}\mathbf{w}_k\|^2+N_T\sigma_n^2},
\end{align}
where
\begin{align}
\mathbf{a}^H(\theta_E,R_E)\mathbf{a}(\theta_D,R_D)=\sum_{n=1}^{N_T}e^{-j2\pi\Delta\theta_n},\\
\mathbf{a}^H(\theta_D,R_D)\mathbf{a}(\theta_E,R_E)=\sum_{n=1}^{N_T}e^{j2\pi\Delta\theta_n},
\end{align}
with
\begin{align}
&\Delta\theta_n\\ \nonumber
&=\Delta f_{\eta(n)}\frac{R_E-R_D}{c}-\frac{1}{2}\left(\cos\theta_E-\cos\theta_D\right)(n-1),
\end{align}
and
\begin{align}
&\|\mathbf{a}^H(\theta_E,R_E)\mathbf{T}\mathbf{w}_k\|^2=\mathbf{a}^H(\theta_E,R_E)\mathbf{T}\mathbf{a}(\theta_E,R_E)\\ \nonumber
&=N_T-\frac{1}{N_T}\mathbf{a}^H(\theta_E,R_E)\mathbf{a}(\theta_D.R_D)\mathbf{a}^H(\theta_D,R_D)\mathbf{a}(\theta_E,R_E).
\end{align}
Let us define
\begin{align}
\lambda=\frac{1}{N_T}\sum_{n=1}^{N_T}e^{-j2\pi\Delta\theta_n}\cdot\frac{1}{N_T}\sum_{n=1}^{N_T}e^{j2\pi\Delta\theta_n}.
\end{align}
which ensures that $0\leq\lambda<1$.
Then,
\begin{align}\label{SINR_E}
\text{SINR}(\theta_E,R_E)=\frac{\mu_1\lambda}{\mu_2(1-\lambda)+\mu_3},
\end{align}
where $\mu_1=\rho^2P_S\beta_1^2$, $\mu_2=\rho^2P_S\beta_2^2$ and $\mu_3=\sigma_n^2$. Obviously, the above function $\text{SINR}(\theta_E,R_E)$ is an increasing function of independent variable $\lambda$ for given fixed $\mu_1=\rho^2P_S\beta_1^2$, $\mu_2=\rho^2P_S\beta_2^2$ and $\mu_3=\sigma_n^2$. As $\lambda$ varies from 0 to 1, the  range of  this function is
\begin{align}
\left[\frac{\mu_1\lambda}{\mu_2+\mu_3},~\frac{\mu_1\lambda}{\mu_3}\right].
\end{align}
where $\lambda=1$ implies that the receiver locates at the desired position.  In what follows, we will determine the maximum value of $\lambda$  when the eavesdropper lies in wiretap region. It is certain that the maximum value of $\lambda$ is smaller than one since the eavesdropper is assumed to be outside the main-beam around the desired position $(\theta_D,R_D)$. Here, we define the wiretap region as all region outside main-peak of SINR around the desired position $(\theta_D,R_D)$. Now, to make the wiretap region approximately clear, we should compute the first nulls of SINR along phase and distance dimensions. In such positions, $\lambda=0$. In other words, the lowest value of SINR is reached.  The null points of array pattern along the direction dimension satisfy the following condition \cite{Kraus}
\begin{align}
&\sum_{n=1}^{N_T}e^{-j2\pi\Delta\theta_n}=\sum_{n=1}^{N_T}e^{j2\pi\frac{1}{2}\left(\cos\theta_E-\cos\theta_D\right)(n-1)}\\ \nonumber
&~~~~=\frac{1-e^{j\pi(\cos\theta_E-\cos\theta_D)N_T}}{1-e^{j\pi(\cos\theta_E-\cos\theta_D)}}=0,
\end{align}
which yields
\begin{align}
&N_T\pi(\cos\theta_E-\cos\theta_D)=\pm 2K\pi,\\ \nonumber
&~~~~K\neq mN_T~(m=1,2,3,...).
\end{align}
In the same manner, the null points along the distance dimension are
\begin{align}
&2\pi N_T\frac{B}{N_T}\frac{R_E-R_D}{c}=\pm2K\pi,\\ \nonumber
&~~~~K\neq mN_T~(m=1,2,3,...).
\end{align}
When $K=1$, we will get the first-null position
\begin{align}
&\theta_{E_{\mathrm{zero}}}=\arccos(\cos\theta_D\pm\frac{2}{N_T}),\\
&R_{E_{\mathrm{zero}}}=R_D\pm\frac{c}{B}.
\end{align}
Then, $\Delta\theta=|\theta_{E_{\mathrm{zero}}}-\theta_D|$ and $\Delta R=|R_{E_{\mathrm{zero}}}-R_D|$.
While $\lambda=\lambda_{\max}$, we can get the maximum value of $\text{SINR}(\theta_E,R_E)$ in the eavesdropper region. Hence, once the eavesdropper located at the peak position of the first side lobe, $\lambda$ will achieve the maximum value. The maximum value of the sidelobe falls between the first and  second nulls. As for the direction dimension, the maximum values of sidelobes are approximately equivalent to
\begin{align}
\sin(\pi(\cos\theta_E-\cos\theta_D)N_T/2)=1.
\end{align}
which yields
\begin{align}
&\pi(\cos\theta_E-\cos\theta_D)\frac{N_T}{2}=\pm(2K+1)\frac{\pi}{2},~(K=1,2,3...).
\end{align}
Similar to the above analysis,  the  condition associated with the distance dimension satisfies
\begin{align}
&2\pi\frac{B}{N_T}\frac{R_E-R_D}{c}\frac{N_T}{2}=\pm(2K+1)\frac{\pi}{2},~(K=1,2,3...).
\end{align}
When $K=1$, we will get two maximum values corresponding to two first sidelobes along phase and distance dimensions as follows
\begin{align}
&\theta_{E_{\mathrm{sidelobe}}}=\arccos(\cos\theta_D\pm\frac{3}{N_T}),\\
&R_{E_{\mathrm{sidelobe}}}=R_D\pm\frac{3c}{2B},
\end{align}
which yields
\begin{align}
\lambda_{\max1}=&\frac{1}{N_T}\sum_{n=1}^{N_T}e^{-j2\pi(\Delta f_{\eta(n)}\frac{R_{E_{\mathrm{sidelobe}}}-R_D}{c})}\\ \nonumber
&\cdot\frac{1}{N_T}\sum_{n=1}^{N_T}e^{j2\pi(\Delta f_{\eta(n)}\frac{R_{E_{\mathrm{sidelobe}}}-R_D}{c})},\\
\lambda_{\max2}=&\frac{1}{N_T}\sum_{n=1}^{N_T}e^{j2\pi(\frac{1}{2}\left(\cos\theta_{E_{\mathrm{sidelobe}}}-\cos\theta_D\right)(n-1))}\\ \nonumber
&\cdot\frac{1}{N_T}\sum_{n=1}^{N_T}e^{-j2\pi(\frac{1}{2}\left(\cos\theta_{E_{\mathrm{sidelobe}}}-\cos\theta_D\right)(n-1))}.
\end{align}
Hence,
\begin{align}
&\text{SINR}(\theta_E,R_E)\\ \nonumber
&~~~~\leq\max\left\{\text{SINR}(\lambda_{\max1}),~\text{SINR}(\lambda_{\max2})\right\}.
\end{align}
This completes the proof of the tight upper bound.
\hfill$\blacksquare$\label{Append_B}

\ifCLASSOPTIONcaptionsoff
\newpage
\fi
\bibliographystyle{IEEEtran}

\bibliography{IEEEfull,REF}

\newpage
\end{document}